\documentclass[11pt,a4paper]{article}

\usepackage{graphicx}%
\usepackage{multirow}%
\usepackage{amsmath,amssymb,amsfonts}%
\usepackage{amsthm}%
\usepackage{mathrsfs}%
\usepackage[title]{appendix}%
\usepackage{xcolor}%
\usepackage{textcomp}%
\usepackage{manyfoot}%
\usepackage{booktabs}%
\usepackage{algorithm}%
\usepackage{algorithmicx}%
\usepackage{algpseudocode}%
\usepackage{listings}%
\usepackage{calligra}

\usepackage{multirow}
\usepackage[mathscr]{eucal}
\usepackage{mathtools}

\usepackage{array}
\usepackage{mdwtab}
\usepackage{tabularray}

\usepackage{float}
\usepackage{stfloats}

\usepackage{tikz}
\usepackage{circuitikz}
\usetikzlibrary{arrows,shapes,backgrounds,circuits.ee.IEC,circuits.logic.US}
\usetikzlibrary {calc,positioning,shapes.misc}
\usetikzlibrary{%
	arrows,%
	shapes.misc,
	shapes.arrows,%
	chains,%
	matrix,%
	positioning,
	scopes,%
	decorations.pathmorphing,
	shadows%
}

\newcommand{\Z}{\mathbb{Z}}
\newcommand{\Q}{\mathbb{Q}}
\newcommand{\C}{\mathbb{C}}

\newcommand{\R}{\mathbb{R}}

\newcommand{\F}{\mathbb{F}}
\newcommand{\E}{\mathbb{E}}

\DeclareMathOperator{\vol}{vol}

\DeclareMathOperator{\Nm}{N}

\newtheorem{Def}{Definition}[]
\newtheorem{Prop}{Proposition}[]
\newtheorem{Teo}{Theorem}[]
\newtheorem{Cor}{Corollary}
\newtheorem{remark}{Remark}[]
\newtheorem{Lema}{Lemma}[]

\title{Construction $\pi_A$ over Multiquadratic Fields for Compound Block-Fading Wiretap Channels}

\usepackage{authblk}
\author[1]{Juliana Souza}
\author[2]{Conghui Li}
\author[3]{Cong Ling}
\affil[1]{IMECC, University of Campinas \thanks{julianagfs@ime.unicamp.br}}
\affil[2,3]{Department of Electrical and Electronic Engineering, Imperial College London\thanks{\{conghui.li15, c.ling\}@imperial.ac.uk}}

\begin{document}

\maketitle

\begin{abstract}
We construct multilevel lattice codes from multiquadratic number fields for the compound block-fading wiretap channel. More precisely, we specialize Construction $\pi_A$ over the ring of integers $\mathcal{O}_K$ and exploit rational primes that split completely in $K$ to obtain a Chinese Remainder Theorem (CRT) decomposition into small residue alphabets, notably binary, which enables multistage decoding. The resulting nested lattices fit into the algebraic Construction~A framework and, when combined with discrete Gaussian shaping and flatness-factor bounds, provide universal reliability for the legitimate receiver and strong secrecy uniformly over the eavesdropper compound set.
\end{abstract}

\medskip


\medskip

\section{Introduction}

Lattice codes are finite constellations obtained by intersecting a lattice with a bounded shaping region. They can approach the capacity of the additive white Gaussian noise (AWGN) channel and support algebraic operations such as compute-and-forward and distributed source coding \cite{Nam2010two-relay, krithivasan2007distributed, zamir2014lattice}.

Lattices also provide a natural framework for physical-layer security. In a wiretap channel, a legitimate receiver and an eavesdropper observe noisy versions of the same transmission \cite{wyner1975wire, leung1978gaussian}. Nested lattice codes combined with discrete Gaussian shaping and flatness-factor analysis can achieve strong secrecy \cite{ling2014semantically}. This approach has been extended to algebraic Construction~A lattices over number fields and to compound multiple-input multiple-output (MIMO) models, yielding universal reliability and flatness-factor-based secrecy guarantees \cite{campello2018universal, campello2020semantically}.

From a practical viewpoint, multilevel lattice constructions such as Construction~$\pi_A$ are appealing because they combine small-alphabet component codes through the CRT and admit multistage decoding, thereby reducing decoding complexity \cite{huang2017construction}. Variants over number fields and quaternionic orders have shown good reliability \cite{huang2017construction, huang2018layered, jsouza2024multilevel}. However, to the best of our knowledge, the physical-layer security of Construction~$\pi_A$ for compound block-fading wiretap channels has not yet been studied.

In this work, we adapt Construction~$\pi_A$ to the ring of integers $\mathcal{O}_K$ of multiquadratic number fields $K$, focusing on rational primes that split completely. This produces a CRT decomposition into small residue alphabets, notably binary ones, and hence enables multistage decoding. We illustrate the construction on $K=\mathbb{Q}(\sqrt{17},\sqrt{33})$, where $\langle 2\rangle$ splits into four coprime ideals. The resulting nested lattice pairs fit into the algebraic Construction~A framework of \cite{campello2018universal, campello2020semantically}, which allows us to obtain universal reliability for Bob and strong secrecy for the corresponding compound block-fading wiretap channel.

Section~2 recalls the algebraic preliminaries on multiquadratic number fields. Section~3 develops Construction~$\pi_A$ over $\mathcal{O}_K$ and presents a binary example over $K=\mathbb{Q}(\sqrt{17},\sqrt{33})$. Section~4 applies the resulting nested lattices to a compound block-fading wiretap channel and includes a finite-length simulation. Finally, Section~5 concludes.

\section{Preliminaries}

We recall the algebraic number theory ingredients needed for our lattice construction; see, e.g., \cite{marcus1977number,jarvis2014algebraic}.

A number field $K$ is a finite extension of $\Q$ of degree $m=[K:\Q]$. Its ring of integers $\mathcal O_K$ is a free $\Z$-module of rank $m$. For an ideal $\mathfrak a\subseteq \mathcal O_K$, its norm is \(\Nm(\mathfrak a)=|\mathcal O_K/\mathfrak a|.\)

The CRT is fundamental to our construction: if $\mathfrak p_1,\ldots,\mathfrak p_r$ are pairwise coprime ideals and \( \mathfrak P=\prod_{j=1}^r \mathfrak p_j, \) then
\[
\mathcal O_K/\mathfrak P \cong (\mathcal O_K/\mathfrak p_1)\times\cdots\times(\mathcal O_K/\mathfrak p_r).
\]

\begin{Def}
A biquadratic field is a field of the form \( K=\Q(\sqrt a,\sqrt b), \)
where $a,b$ are distinct square-free integers. Then $[K:\Q]=4$, and $K$ contains the three quadratic subfields $\Q(\sqrt a)$, $\Q(\sqrt b)$, and $\Q(\sqrt k)$, where $k=ab/\gcd(a,b)^2$.
\end{Def}

For a biquadratic field $K=\Q(\sqrt a,\sqrt b)$, the ring of integers $\mathcal O_K$ admits an explicit integral basis depending on the congruence classes of $a,b,$ and $k$ modulo $4$; see \cite{jarvis2014algebraic}.

A rational prime $p$ factors in $\mathcal O_K$ as \( p\mathcal O_K=\prod_{i=1}^r \mathfrak p_i^{e_i}, \) where $f_i=[\mathcal O_K:\F_p]$ and $\sum_i e_i f_i = m$. We say that $p$ splits completely if $r=m$ and $e_i=f_i=1$ for all $i$.

For a quadratic field $\Q(\sqrt a)$ and an odd prime $p\nmid a$, the Legendre symbol determines the splitting behaviour:
\[
\Big(\frac{a}{p}\Big)=1 \Rightarrow p \text{ splits},\qquad
\Big(\frac{a}{p}\Big)=-1 \Rightarrow p \text{ is inert}.
\]
If $p\mid a$, then $p$ ramifies. For $p=2$, the usual congruence criteria apply.

\begin{Prop}[\cite{marcus1977number}]\label{completelysplits}
If $p$ splits completely in both $\Q(\sqrt a)$ and $\Q(\sqrt b)$, then it splits completely in $\Q(\sqrt a,\sqrt b)$.
\end{Prop}

Therefore, when a rational prime splits completely in a biquadratic field, the CRT yields a decomposition into several small residue fields. This is the key algebraic ingredient used in the next section to define Construction $\pi_A$ over $\mathcal O_K$ and to enable multistage decoding.

\section{Lattice Constructions from Multiquadratic Fields}

In this section, we recall algebraic lattices via the canonical embedding and define Construction~$\pi_A$ over $\mathcal O_K$ using the CRT, leading to nested lattice pairs with explicit design rate.

\subsection{Algebraic Lattices and Canonical Embeddings}

A lattice in $\R^N$ is a discrete subgroup generated by $N$ linearly independent vectors. If $G=[\mathbf v_1|\cdots|\mathbf v_N]$ is a generator matrix, then
\[
\Lambda=\{G\mathbf u:\mathbf u\in\Z^N\}.
\]

Let $K$ be a number field of degree $m=[K:\Q]$ and signature $(r_1,r_2)$. The canonical embedding $\sigma_K:K\hookrightarrow \R^{r_1}\times\C^{r_2}\simeq\R^m$ maps $\mathcal O_K$ to a full-rank lattice in $\R^m$. In the biquadratic case $K=\Q(\sqrt a,\sqrt b)$, we have $m=4$; in particular, if $a,b>0$, then $K$ is totally real and $(r_1,r_2)=(4,0)$.

More generally, if $\{x_1,\dots,x_m\}$ is a $\Z$-basis of a $\Z$-module $M\subset K$, then
\[
\sigma_K(M)=\left\{\sum_{i=1}^m u_i\,\sigma_K(x_i)\;\middle|\;u_i\in\Z\right\}
\]
is a lattice in $\R^m$.

In our setting, lattice points lie in $\mathcal O_K^n$, where $n$ denotes the code length over $\mathcal O_K$. Applying the canonical embedding componentwise, we obtain Euclidean lattices in $\R^N$, where $N=mn$. By abuse of notation, we still denote this componentwise embedding by $\sigma_K:\mathcal O_K^n\hookrightarrow \R^N.$

The volume of the lattice associated with $\mathcal O_K$ is \( \vol(\sigma_K(\mathcal O_K))=2^{-r_2}\sqrt{|d_K|}, \) where $d_K$ denotes the discriminant of $K$. More generally, for an ideal $\mathfrak a\subseteq \mathcal O_K$, \( \vol(\sigma_K(\mathfrak a))=2^{-r_2}\sqrt{|d_K|}\,\Nm(\mathfrak a), \) \cite{samuel2013algebraic}.


\subsection{Construction $\pi_A$ Lattices}

We now adapt Construction $A$ to a multilevel setting using the CRT, following Construction $\pi_A$ in \cite{huang2017construction}.%
\footnote{For a linear code $C\subseteq \mathbb Z_p^N$, Construction $A$ defines the lattice
$\Lambda_A(C)=\{x\in\mathbb Z^N: x \bmod p \in C\}$.
In the algebraic setting, $\mathbb Z$ and $\mathbb Z_p$ are replaced by $\mathcal O_K$ and $\mathcal O_K/\mathfrak p$.}

Let $\mathfrak p_1,\dots,\mathfrak p_r\subset \mathcal O_K$ be pairwise coprime prime ideals, and set $\mathfrak P=\prod_{j=1}^r \mathfrak p_j.$ Let \( \rho:\mathcal O_K^n \to (\mathcal O_K/\mathfrak P)^n \)
denote the componentwise reduction modulo $\mathfrak P$, and let \( \phi:\mathcal O_K/\mathfrak P \rightarrow
\mathcal O_K/\mathfrak p_1\times\cdots\times\mathcal O_K/\mathfrak p_r \) be the CRT isomorphism.

\begin{Def}[Construction $\pi_A$ lattice]\label{def:cons_piA}
Let $\mathcal C_j\subseteq (\mathcal O_K/\mathfrak p_j)^n$ be $(n,k_j)$-linear codes for $j=1,\dots,r$, and define
\[
\mathcal C:=\phi^{-1}(\mathcal C_1\times\cdots\times\mathcal C_r)\subseteq (\mathcal O_K/\mathfrak P)^n.
\]
The corresponding Construction $\pi_A$ lattice is $\Lambda_{\pi_A}(\mathcal C):=\rho^{-1}(\mathcal C)\subseteq \mathcal O_K^n.$ Equivalently, $\Lambda_{\pi_A}(\mathcal C)=\mathcal C+\mathfrak P^n.$
\end{Def}

This construction enables multistage decoding by decomposing the lattice decoding into smaller component decoding tasks over the residue fields $\mathcal O_K/\mathfrak p_j$. Its volume is $\vol(\Lambda_{\pi_A}(\mathcal C)) = \prod_{j=1}^r \Nm(\mathfrak p_j)^{\,n-k_j}\, 2^{-nr_2}|d_K|^{n/2}.$

\subsubsection{Binary Example}

We illustrate the construction when the rational prime $2$ splits completely, so that the residue alphabets are binary.

First consider $K=\Q(\sqrt{17})$. Writing $\omega=\frac{1+\sqrt{17}}{2}$, the ideal $\langle 2\rangle$ splits as $\langle 2\rangle=\mathfrak p_1\mathfrak p_2.$ Hence, by the CRT, $\mathcal O_K/\langle 2\rangle \cong \mathcal O_K/\mathfrak p_1 \times \mathcal O_K/\mathfrak p_2 \cong \F_2\times \F_2. $ A convenient isomorphism is
\begin{align*}
\Psi:\mathcal O_K/\langle 2\rangle &\to \F_2\times\F_2,\\
\alpha &\mapsto (\alpha \bmod \mathfrak p_1,\alpha \bmod \mathfrak p_2).    
\end{align*}

Now consider the biquadratic field $K=\Q(\sqrt{17},\sqrt{33}).$ Since $\langle 2\rangle$ splits completely in both $\Q(\sqrt{17})$ and $\Q(\sqrt{33})$, Proposition~\ref{completelysplits} implies that it also splits completely in $K$: \( \langle 2\rangle=\mathfrak p_1\mathfrak p_2\mathfrak p_3\mathfrak p_4. \) The explicit prime ideals can be computed in SageMath, \cite{sagemath}.
Therefore,
\[
\mathcal O_K/\langle 2\rangle
\cong
\mathcal O_K/\mathfrak p_1\times\cdots\times \mathcal O_K/\mathfrak p_4
\cong
\F_2^4.
\]

Thus Construction $\pi_A$ over $K$ can be built from four binary component codes, substantially reducing decoding complexity while preserving the algebraic structure needed for nested lattice coding.

\begin{remark}
Construction $\pi_A$ admits multistage decoding strategies such as the serial modulo decoder (SMD) and the parallel modulo decoder (PMD) from \cite{huang2017construction}. SMD decodes the CRT levels successively, whereas PMD decodes them in parallel. Both have comparable arithmetic complexity per codeword, while PMD may offer lower latency on parallel hardware at a modest performance loss.
\end{remark}

\subsection{Nested Lattice Codes from Construction $\pi_A$ over Multiquadratic Fields}\label{sec:nested-piA}

We now specialize Construction $\pi_A$ to build nested lattice pairs for the wiretap setting. Let $K$ be a multiquadratic field of degree $m=[K:\Q]$ with ring of integers $\mathcal O_K$. Let $\mathfrak p_1,\dots,\mathfrak p_r\subset \mathcal O_K$ be pairwise coprime ideals, with $q_j=\Nm(\mathfrak p_j), \mathfrak P=\prod_{j=1}^r \mathfrak p_j.$

For each level $j\in\{1,\dots,r\}$, choose nested linear codes \( \mathcal C_e^{(j)} \subseteq \mathcal C_b^{(j)} \subseteq (\mathcal O_K/\mathfrak p_j)^n, \) with dimensions $k_{e,j}\le k_{b,j}$. Using the CRT isomorphism from Section~3.2, define the global nested codes
\begin{align*}
\mathcal C_e&:=\phi^{-1}\Big(\mathcal C_e^{(1)}\times\cdots\times \mathcal C_e^{(r)}\Big)
\subseteq (\mathcal O_K/\mathfrak P)^n,\\
\mathcal C_b&:=\phi^{-1}\Big(\mathcal C_b^{(1)}\times\cdots\times \mathcal C_b^{(r)}\Big)
\subseteq (\mathcal O_K/\mathfrak P)^n.
\end{align*}

Applying the Construction $\pi_A$ lifting from Definition~\ref{def:cons_piA}, we obtain
\[
\widetilde{\Lambda}_e=\mathcal C_e+\mathfrak P^n, \qquad
\widetilde{\Lambda}_b=\mathcal C_b+\mathfrak P^n,
\]
with $\widetilde{\Lambda}_e\subseteq \widetilde{\Lambda}_b\subseteq \mathcal O_K^n$.

Using the canonical embedding $\sigma_K:\mathcal O_K^n\hookrightarrow \R^{N}$ and a scaling factor $\gamma>0$, we obtain the Euclidean nested lattice pair
\[
\Lambda_e:=\gamma\sigma_K(\widetilde{\Lambda}_e), \qquad
\Lambda_b:=\gamma\sigma_K(\widetilde{\Lambda}_b),
\]
with $\Lambda_e\subseteq \Lambda_b\subset \R^{N}$. Here $\Lambda_b$ will serve as Bob's fine lattice, whereas $\Lambda_e$ will be the coarse lattice used for shaping and secrecy.

The quotient size is \( |\Lambda_b/\Lambda_e| = \frac{\vol(\Lambda_e)}{\vol(\Lambda_b)} = \prod_{j=1}^r q_j^{\,k_{b,j}-k_{e,j}}, \) and therefore the design rate is  \( R_{\mathrm{design}} = \frac{1}{N}\log_2 |\Lambda_b/\Lambda_e| = \frac{1}{mn}\sum_{j=1}^r (k_{b,j}-k_{e,j})\log_2 q_j.\)
The scaling $\gamma$ is chosen to satisfy the transmit power constraint.

\begin{remark}
The simplest multistage decoding arises when the ideals $\mathfrak p_j$ lie above rational primes that split completely in $K$. In that case, \( \mathcal O_K/\mathfrak p_j \cong \F_{p_j}, \) so each level is defined over a small finite field. More generally, the construction applies to any collection of pairwise coprime ideals, with residue field sizes $q_j=\Nm(\mathfrak p_j)$ not necessarily prime.
\end{remark}

\section{Wiretap Application: Compound Block Fading}

We now apply the nested lattices from multiquadratic Construction~$\pi_A$ to a compound block-fading wiretap channel. The nested pair $\Lambda_e\subset\Lambda_b\subset\R^N$ plays two distinct roles: the fine lattice $\Lambda_b$ is used to guarantee reliable decoding at Bob, whereas the coarse lattice $\Lambda_e$ is used for shaping and for hiding the transmitted coset from Eve. Our approach combines the universality result of~\cite{campello2018universal} with the flatness-factor based secrecy framework of~\cite{ling2014semantically,campello2020semantically}.

\subsection{Channel Model and Compound Sets}

As in Section~3.3, let $\Lambda_e \subset \Lambda_b \subset \R^N$ be a nested lattice pair obtained from multiquadratic Construction~$\pi_A$. In this section, rates are measured in nats per channel use.

We consider a quasi-static compound block-fading complex wiretap channel with $n_a$ transmit antennas, $n_b$ antennas at Bob, and $n_e$ antennas at Eve. During one codeword of $T$ channel uses,
\[
Y_b = H_b X + Z_b,\qquad Y_e = H_e X + Z_e,
\]
where $H_b\in\C^{n_b\times n_a}$ and $H_e\in\C^{n_e\times n_a}$ are the channel matrices, $X\in\C^{n_a\times T}$ is the transmitted codeword, and $Z_b,Z_e$ have i.i.d.\ entries distributed as $\mathcal{CN}(0,\sigma_b^2)$ and $\mathcal{CN}(0,\sigma_e^2)$, respectively. We impose the average power constraint 
$\frac{1}{n_aT}\E[\|X\|_F^2]\le P,$ and define $\mathrm{SNR}_b=\frac{P}{\sigma_b^2}$ and $\mathrm{SNR}_e=\frac{P}{\sigma_e^2}.$

Following~\cite{campello2020semantically}, we define the admissible channel families
\begin{align}
S_b &:= \Big\{ H\in\C^{n_b\times n_a} :
     \big| I_{n_a} + \mathrm{SNR}_b H^\dagger H \big| \ge e^{C_b} \Big\}, \\
S_e &:= \Big\{ H\in\C^{n_e\times n_a} :
     \big| I_{n_a} + \mathrm{SNR}_e H^\dagger H \big| \le e^{C_e} \Big\}.
\end{align}
A channel realization is admissible if $H_b\in S_b$ and $H_e\in S_e$. Thus $S_b$ collects the channel realizations for which Bob’s channel is sufficiently good, while $S_e$ collects those for which Eve’s channel is no stronger than the prescribed threshold. The corresponding secrecy-capacity lower bound is $C_s=(C_b-C_e)^+,$ and our goal is to achieve any secrecy rate $R<C_s$ uniformly over all admissible pairs $(H_b,H_e)\in S_b\times S_e$ (up to the constant gap inherited from the reduction step).

Vectorizing the channel gives
\[
\mathrm{vec}(Y_b)=(I_T\otimes H_b)\mathrm{vec}(X)+\mathrm{vec}(Z_b),
\quad
\mathrm{vec}(Y_e)=(I_T\otimes H_e)\mathrm{vec}(X)+\mathrm{vec}(Z_e).
\]
After passing to the equivalent real model induced by the lattice embedding, we obtain \(\mathbf y_b = H_b^{\mathrm{eq}}\mathbf x_R + \mathbf z_b,\) and \( \mathbf y_e = H_e^{\mathrm{eq}}\mathbf x_R + \mathbf z_e,
\) where $\mathbf x_R\in\R^N$ corresponds to a point in $\Lambda_b$.

\subsection{Reliability}

We now show that the fine lattices from multiquadratic Construction~$\pi_A$ are universally good for Bob’s compound channel family $S_b$, meaning that they remain decodable for every admissible realization in $S_b$. The key observation is that our CRT-based multilevel construction is a sub-ensemble of the algebraic Construction~A lattices over $\mathcal O_K$ studied in~\cite{campello2018universal,campello2020semantically}.

Fix a code length $n$ and pairwise coprime prime ideals $\mathfrak p_1,\dots,\mathfrak p_r\subset \mathcal O_K$, and set \( \mathfrak P:=\prod_{j=1}^r \mathfrak p_j. \) At each level $j$, choose a linear code
\( \mathcal C_j\subset (\mathcal O_K/\mathfrak p_j)^n \) from a balanced ensemble in the sense of~\cite{campello2018random}, and define the CRT code \( \mathcal C_b:=\phi^{-1}(\mathcal C_1\times\cdots\times\mathcal C_r)\subset (\mathcal O_K/\mathfrak P)^n.\)
The corresponding fine lattice $\Lambda_b$ is therefore an algebraic Construction~A lattice over $\mathcal O_K$ with modulus $\mathfrak P$.

Let $\mathbb L_n$ denote the ensemble obtained by drawing the component codes independently from balanced ensembles and forming the corresponding lattices $\Lambda_b$. By~\cite{campello2020random,jsouza2024multilevel}, Construction~$\pi_A$ yields a balanced ensemble, and hence inherits the same averaging properties needed in~\cite{campello2018universal}.

\begin{Prop}[Minkowski--Hlawka and compactification]
\label{prop:MH+comp-piA}
For each $n$, the multiquadratic Construction~$\pi_A$ ensemble $\mathbb L_n$ satisfies: (i)\emph{(Minkowski--Hlawka)} For every non-negative continuous function $f$ with compact support,
\[
\mathbb E_{\Lambda_b\in\mathbb L_n}\!\left[\sum_{\lambda\in\Lambda_b\setminus\{0\}} f(\lambda)\right]
\le
\frac{1}{\vol(\Lambda_b)}\int_{\R^N} f(\mathbf u)\,d\mathbf u + \varepsilon_n,
\]
with $\varepsilon_n\to 0$ as $n\to\infty$. (ii) \emph{(Compactification)} The family $\{\mathbb L_n\}_{n\ge 1}$ compactifies the normalized block-fading channel family in the sense of~\cite{campello2018universal}: for every normalized channel matrix $\widetilde H$ with $\det(\widetilde H^\dagger \widetilde H)=1$, one can write \( \widetilde H = E_{\widetilde H}U_{\widetilde H},\) where $\|E_{\widetilde H}\|$ is uniformly bounded and $U_{\widetilde H}$ preserves the distribution of the ensemble.
\end{Prop}

\emph{Sketch of proof.}
Item~(i) follows from the balancedness of the CRT code ensemble over $(\mathcal O_K/\mathfrak P)^n$~\cite{campello2018random,jsouza2024multilevel} together with the standard Minkowski--Hlawka averaging argument over the base lattice $\sigma_K(\mathcal O_K^n)$ as in~\cite[Sec.~III]{campello2018universal}. Item~(ii) follows from the same algebraic reduction argument used in~\cite{campello2018universal}: Dirichlet’s unit theorem yields a full-rank lattice of logarithmic embeddings of units in $\mathcal O_K$, and a suitable unit gives the factorization $\widetilde H = E_{\widetilde H}U_{\widetilde H}$ with bounded $\|E_{\widetilde H}\|$.

\begin{Cor}[Universal reliability at Bob]
\label{cor:reliability-compound}
Let $\{\mathbb L_n\}$ be the multiquadratic Construction~$\pi_A$ ensembles above, with fixed volume and increasing dimension $N$. Then there exists a sequence of lattices $\Lambda_b^{(n)}\in\mathbb L_n$ that is universally good for Bob’s compound channel family $S_b$, in the sense of~\cite{campello2018universal}.
\end{Cor}

This is an immediate consequence of~\cite[Theorem~1]{campello2018universal}.

\subsection{Secrecy via Flatness Factor and Algebraic Reduction}

We now turn to secrecy. Intuitively, secrecy requires that Eve’s received distribution be essentially independent of the transmitted message, so that different cosets become indistinguishable at the output.

Our goal is to obtain universal strong secrecy over Eve’s compound set, namely
\[
\sup_{H_e\in S_e} I(M;Y_e)\to 0
\qquad\text{as } n\to\infty.
\]

The secrecy argument follows~\cite{ling2014semantically,campello2020semantically} and has three steps: (i) discrete Gaussian shaping randomizes the transmitted point inside each coset and reduces secrecy to a flatness-factor bound; (ii) algebraic reduction replaces the channel-dependent flatness factor by a spherical one at an equivalent variance $\sigma_{\mathrm{eq}}$; and (iii) the Minkowski--Hlawka property guarantees a sequence of lattices for which this spherical flatness factor vanishes.

We use a discrete Gaussian encoder over the coarse lattice $\Lambda_e$. For each message $s\in M_n$, Alice samples a lattice point in the coset $\Lambda_e+\lambda_s$ according to the discrete Gaussian distribution $D_{\Lambda_e+\lambda_s,\sigma_s}$ and transmits the corresponding codeword. Let $p_{Y_e|M=s}$ denote Eve’s output density given $M=s$. If there exist a reference density $q_{Y_e}$ and a sequence $\varepsilon_n\in[0,1/4]$ such that
\(\mathbb V(p_{Y_e|M=s},q_{Y_e})\le \varepsilon_n\) for all \(s\in M_n, \) then~\cite[Lemma~2]{campello2020semantically} implies
\[
I(M;Y_e)\le 8n\varepsilon_n R - 8\varepsilon_n\log_2(8\varepsilon_n).
\]
Hence it is enough to prove that $\varepsilon_n=o(1/n)$ uniformly over $H_e\in S_e$.

Fix an admissible eavesdropper realization $H_e\in S_e$. As shown in~\cite{campello2020semantically}, the variational-distance parameter can be bounded by a flatness factor of the lattice $H_e\Lambda_e$ with respect to a correlated Gaussian
\[
\varepsilon_n \le \epsilon_{H_e\Lambda_e}\big(\sqrt{\Sigma_3}\big),
\qquad
\Sigma_3^{-1}=(H_eH_e^\dagger)^{-1}\sigma_s^{-2}+\sigma_e^{-2}I.
\]
The difficulty is that this quantity depends on the particular channel realization $H_e$. To obtain a bound that is uniform over the entire compound set, we introduce the worst-case quantity
\[
C_e(H_e):=\log\det\!\Big(I+\frac{\sigma_s^2}{\sigma_e^2}H_eH_e^\dagger\Big),
\qquad
C_e:=\sup_{H_e\in S_e} C_e(H_e).
\]
The next lemma replaces the channel-dependent flatness factor by a spherical one at a channel-independent equivalent variance.

\begin{Lema}[Flatness factor under algebraic reduction, {\cite{campello2020semantically}}]
\label{lem:flatness-reduction}
Assume that the dual lattice $\Lambda^\ast$ admits algebraic reduction with constant $\alpha>0$. Then, for every $H_e\in S_e$,
\[
\epsilon_{H_e\Lambda}\big(\sqrt{\Sigma_3}\big)\le \epsilon_\Lambda(\sigma_{\mathrm{eq}}),
\qquad
\sigma_{\mathrm{eq}}^2=\sigma_s^2\alpha^{-2}e^{-C_e/n_a},
\]
where $\epsilon_\Lambda(\cdot)$ denotes the spherical flatness factor.
\end{Lema}

In our setting, the coarse lattice $\Lambda_e$ is an algebraic Construction~A lattice over $\mathcal O_K$ obtained from CRT codes. By Proposition~\ref{prop:MH+comp-piA}, the corresponding ensemble satisfies the Minkowski--Hlawka property, and therefore we can choose a sequence $\{\Lambda_e^{(n)}\}$ such that $\epsilon_{\Lambda_e^{(n)}}(\sigma_{\mathrm{eq}})=o(1/n).$ By Lemma~\ref{lem:flatness-reduction}, this yields the uniform estimate
\[
\sup_{H_e\in S_e}\epsilon_{H_e\Lambda_e^{(n)}}\big(\sqrt{\Sigma_3}\big) \le \epsilon_{\Lambda_e^{(n)}}(\sigma_{\mathrm{eq}}) = o(1/n),
\]
and consequently $\sup_{H_e\in S_e} I(M;Y_e)\to 0.$

\begin{Teo} \label{thm:compound-wiretap-main}
Let $K$ be a multiquadratic field and let $\{(\Lambda_b^{(n)},\Lambda_e^{(n)})\}_{n\ge1}$ be nested lattices produced by multiquadratic Construction~$\pi_A$, with $\Lambda_e^{(n)}\subset \Lambda_b^{(n)}$. Assume that:
(i) the fine lattices $\Lambda_b^{(n)}$ are universally good for $S_b$ in the sense of Corollary~\ref{cor:reliability-compound}, and (ii) the coarse lattices $\Lambda_e^{(n)}$ are secrecy-good for AWGN at $\sigma_{\mathrm{eq}}$, and their duals admit algebraic reduction with constant $\alpha$. Then there exists a discrete Gaussian encoder based on $(\Lambda_b^{(n)},\Lambda_e^{(n)})$ and a lattice decoder at Bob such that, for every $(H_b,H_e)\in S_b\times S_e$, Bob’s error probability and the leakage $I(M;Y_e)$ both vanish as $n\to\infty$, provided that the secrecy rate per channel use satisfies
\[
R<\bigl(C_b-C_e-n_a-2n_a\log\alpha\bigr)^+ .
\]
\end{Teo}

\emph{Sketch of proof.}
Reliability follows from Corollary~\ref{cor:reliability-compound}. For secrecy, choose $\Lambda_e^{(n)}$ so that $\epsilon_{\Lambda_e^{(n)}}(\sigma_{\mathrm{eq}})=o(1/n)$ using Proposition~\ref{prop:MH+comp-piA}. Lemma~\ref{lem:flatness-reduction} then gives a uniform bound over all $H_e\in S_e$, and substituting this into the variational-distance estimate above yields strong secrecy. The rate penalty $n_a+2n_a\log\alpha$ is exactly the penalty inherited from the algebraic reduction step in~\cite{campello2020semantically}.

\subsection{Illustrative Finite-Length Example}

We illustrate the proposed construction on $K=\mathbb{Q}(\sqrt{17},\sqrt{33})$, where the complete splitting of $\langle 2\rangle$ yields four binary CRT levels. After embedding and zero-forcing equalization for a fixed $2\times2$ Rayleigh MIMO channel pair $(H_b,H_e)$, the decoder sees equivalent real channels $y_b=x+z_b$ and $y_e=x+z_e,$ with $z_b\sim\mathcal{N}(0,I_N)$ and $z_e\sim\mathcal{N}(0,6I_N)$, consistent with the single-block compound model. Each level uses an independent $(3,6)$-regular LDPC code of length $n=800$ and rate $1/2$, generated by PEG and followed by a random interleaver. We apply multistage decoding with successive interference cancellation and layered belief propagation~\cite{hu2005PEG,tse2005fundamentals,uchoa2016iterative}. Bob’s SNR is swept from $0$ to $24$~dB, and for each point we simulate until about $800$ bit errors are collected at Bob.

\begin{figure}[htpb]
  \centering
  \includegraphics[width=.9\linewidth]{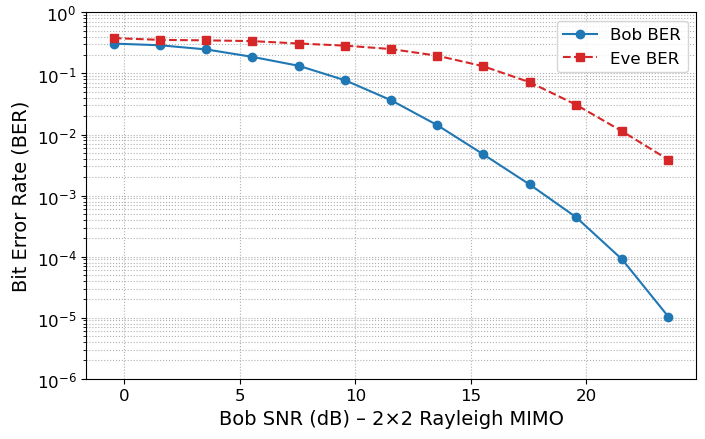}
  \caption{Bit-error rate versus Bob's SNR for the Construction $\pi_A$ over $K=\mathbb{Q}(\sqrt{17},\sqrt{33})$ on a $2\times 2$ Rayleigh MIMO wiretap channel.}
  \label{fig:ber_bob_eve}
\end{figure}

The BER curves in Fig.~\ref{fig:ber_bob_eve} show a clear waterfall for Bob, while Eve remains at a much higher BER due to the effective noise-variance penalty $10\log_{10}6\approx 7.8$~dB.

\section{Conclusions and Perspectives}

We specialized Construction~$\pi_A$ over multiquadratic fields and obtained multilevel lattice codes over $\mathcal{O}_K$ that admit multistage decoding. When a rational prime splits completely, the induced CRT decomposition yields small residue alphabets, notably binary ones, so that standard LDPC component codes can be employed while preserving the algebraic structure of the construction. Placed within the algebraic Construction~A framework, the resulting nested lattices apply to a compound block-fading wiretap channel, where discrete Gaussian shaping and flatness-factor bounds \cite{ling2014semantically,campello2020semantically} yield universal reliability for Bob and strong secrecy against Eve.

Our example over $K=\mathbb{Q}(\sqrt{17},\sqrt{33})$ illustrates the use of binary component codes and the expected finite-length decoding behaviour. As future work, we plan to compare these multiquadratic Construction~$\pi_A$ ensembles with the multilevel algebraic partitions of \cite{campello2017multilevel}, and to develop finite-blocklength reliability and secrecy guarantees based on channel resolvability \cite{polyanskiy2010finite,hayashi2006resolvability,bloch2013strong} for massive Internet-of-Things (IoT) and ultra-reliable low-latency communication (URLLC) scenarios with stringent energy and latency constraints~\cite{wang2019shortpkt}.

\section*{Acknowledgment}
The work of J.G.F. Souza is supported by the São Paulo Research Foundation (FAPESP) under grant 2025/22940-2.

\bibliographystyle{alpha}
\bibliography{bibliography.bib} 

\end{document}